\documentclass[letterpaper]{article}
\usepackage[top=3cm,bottom=3cm,left=3cm,right=3cm,marginparwidth=1.75cm]{geometry}
\usepackage{float}
\usepackage{listings}
\usepackage{xcolor}
\usepackage{multirow}
\usepackage{dcolumn}
\usepackage[english]{babel}
\usepackage[utf8]{inputenc}
\usepackage[T1]{fontenc}
\usepackage{palatino}
\usepackage{amsmath}
\usepackage{afterpage}
\usepackage{graphicx, subcaption}
\usepackage[colorinlistoftodos]{todonotes}
\usepackage[colorlinks=true, linkcolor=blue, citecolor=blue]{hyperref}
\usepackage{fancyhdr}

\begin{document}

\vspace*{1.2cm}

\thispagestyle{empty}
\begin{center}

{\LARGE \bf Semileptonic and Leptonic $\boldsymbol{B}$ Physics at Belle II}

\par\vspace*{7mm}\par

{

\bigskip

\large \bf Priyanka Cheema }

\bigskip

{\large \bf  pche3675@uni.sydney.edu.au}

\bigskip

{University of Sydney}

\bigskip

{\it Presented at the 4th World Summit on Exploring the Dark Side of the Universe \\La Réunion, November 7-11 2022}

\end{center}
 
\begin{abstract}
Semileptonic decays of $B$ mesons that proceed via $b\rightarrow u$ and $b\rightarrow c$ transitions, as well as rare leptonic $B$ decays, are fertile grounds for probing new physics beyond the Standard Model. The Belle II experiment presents as an excellent opportunity to study such decays, and since beginning operation in 2019, Belle II has already recorded half of the data collected by its predecessor and reached an instantaneous luminosity greater than $4\times 10^{34}$ cm$^{-2}$s$^{-1}$. Belle II expects to explore flavour physics with $B$ and charmed mesons with unprecedented precision, and has already produced multiple world-leading results in this area. We present recent results from Belle II on measurements of the CKM parameters $V_{ub}$ and $V_{cb}$ as well as results for $R(X_{e/\mu})$. 
\end{abstract}
  
\section{Belle II and SuperKEKB}
\label{S:1}

The Belle II detector is situated at the interaction point of the SuperKEKB $e^+e^-$ collider located at the KEK laboratory in Tsukuba, Japan. It began operation in March 2019 and has since collected 428 fb$^{-1}$ of data and achieved a world record instantaneous luminosity of 4.7 $\times 10^{34}$ cm$^{-2}$s$^{-1}$. All results presented here use a subset (189 fb$^{-1}$) of the available Belle II data which was collected from 2019-2021. The Belle II detector achieves a near 4$\pi$ solid-angle coverage of the interaction point and consists of cylindrical layers that comprise a magnetic spectrometer, electromagnetic calorimeter and muon detector. A superconducting solenoid provides a 1.5 T magnetic field that envelopes the former two layers. A complete description of the Belle II detector and its sub-detectors is given in \cite{belle2}.  

In addition to measuring key CKM parameters and testing lepton flavour universality, the Belle II physics program also aims to measure the amount of CP asymmetry in the $B$ and $D$ meson systems, perform studies into $b\bar{b}$ and $c\bar{c}$ spectroscopy, search for forbidden $B$ meson decays, and perform dark sector studies. Currently, Belle II and SuperKEKB are in a long shutdown to allow for accelerator and detector upgrades, the most notable being the improvement of the pixel detector. Two additional long shutdowns are planned for the lifetime of Belle II to allow for further upgrades that will aim towards achieving the design luminosity of $6\times 10^{35}$ cm$^{-2}$s$^{-1}$ and a total integrated luminosity of 50 ab$^{-1}$ \cite{b2physics}.  

\subsection{Reconstruction Strategies}
\label{SS:11}

The beam energies at SuperKEKB are tuned to the $\Upsilon$(4S) resonance energy so as to produce an $\Upsilon$(4S) meson at the interaction point which then decays into a pair of $B$ mesons. Following this, the event can be divided into two parts: the tag side which contains one $B$ meson, and the signal side which contains the other $B$ meson and the decay(s) of interest. 

For semileptonic and leptonic analyses, the reconstruction approach can either be tagged or untagged. A tagged analysis involves explicit reconstruction of the $B$ meson in the tag side, and it can be reconstructed either semileptonically or hadronically. Although a hadronic tag allows for a complete understanding of the signal side kinematics (which is especially useful for signal decays involving invisible particles), such a tagging method yields a lower efficiency compared to a semileptonic approach. The converse of a tagged analysis is an untagged analysis whereby the tag side $B$ meson is not reconstructed. This method exhibits the highest reconstruction efficiency but suffers from large backgrounds which are difficult to constrain due to the lack of information on the event kinematics and shape. For a tagged analysis, the tag side $B$ meson is reconstructed using the Full Event Interpretation (FEI) algorithm \cite{FEI}. This supersedes the Full Reconstruction algorithm used at Belle, and uses around 200 boosted decision tree classifiers to reconstruct $\mathcal{O}(10000)$ different $B$ meson decay chains. The result is an algorithm that improves upon the reconstruction efficiency of its Belle counterpart by around 50\% \cite{FEI}. 

For the signal side, the $B$ meson can be reconstructed either in a specific decay mode (exclusively) or in multiple decay modes (inclusively). Both approaches acquire different experimental and theoretical uncertainties but they are complementary and provide powerful cross-checks of each other as their measurements should be in agreement. However, currently there does exist a tension between exclusive and inclusive measurements of the CKM parameters $V_{ub}$ and $V_{cb}$ that is not understood \cite{b2physics}.

\section{\texorpdfstring{$\boldsymbol{V_{ub}}$}{Vub} Results}
\label{S:2}
Obtaining a value for the $V_{ub}$ parameter of the CKM matrix requires studying decays that exhibit $b\rightarrow u$ transitions. Examples of commonly used decays for $V_{ub}$ measurements include the inclusive $B\rightarrow X_u\ell\nu_\ell$ mode, and the exclusive $B\rightarrow \rho\ell\nu_\ell$ and $B\rightarrow \pi\ell\nu_\ell$ modes. Here, a tagged analysis of $B\rightarrow \pi e^+\nu_e$ and an untagged analysis of $B^0\rightarrow\pi^-\ell^+\nu_\ell$ will be presented, while results from the tagged analysis of $B\rightarrow\rho\ell\nu_\ell$ can be found in \cite{rho_analysis}. 

\subsection{Tagged \texorpdfstring{$\boldsymbol{B\rightarrow\pi e^+\nu_e}$}{B → πe+νe}}
\label{SS:21}
\begin{figure}[h]
         \centering
         \begin{subfigure}[t]{0.49\textwidth}
             \centering
             \includegraphics[width=0.9\textwidth]{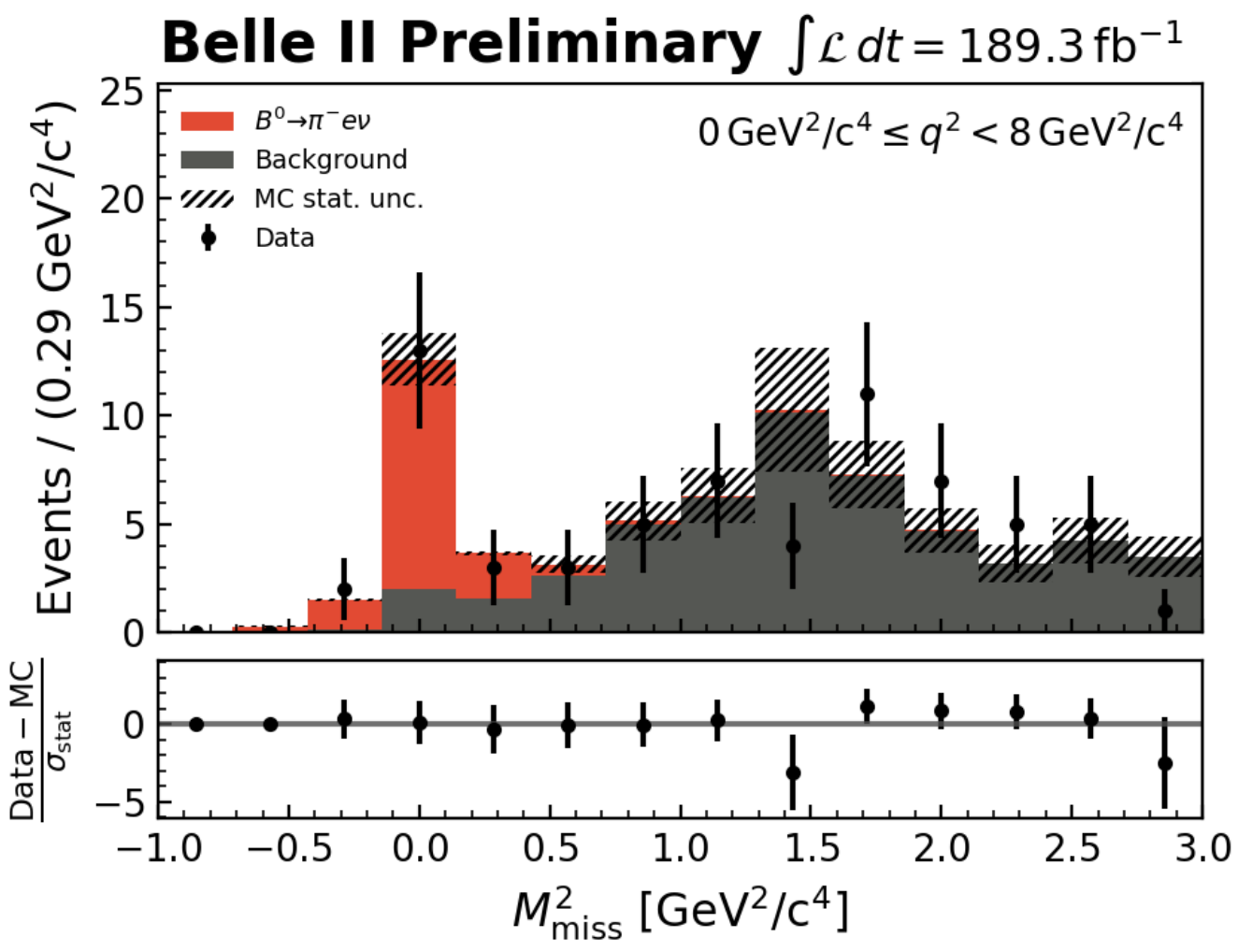}
             \caption{Fitted distribution of $M_\mathrm{miss}^2$ for the first $q^2$ bin}
         \end{subfigure}
         \begin{subfigure}[t]{0.49\textwidth}
             \centering
             \includegraphics[width=0.86\textwidth]{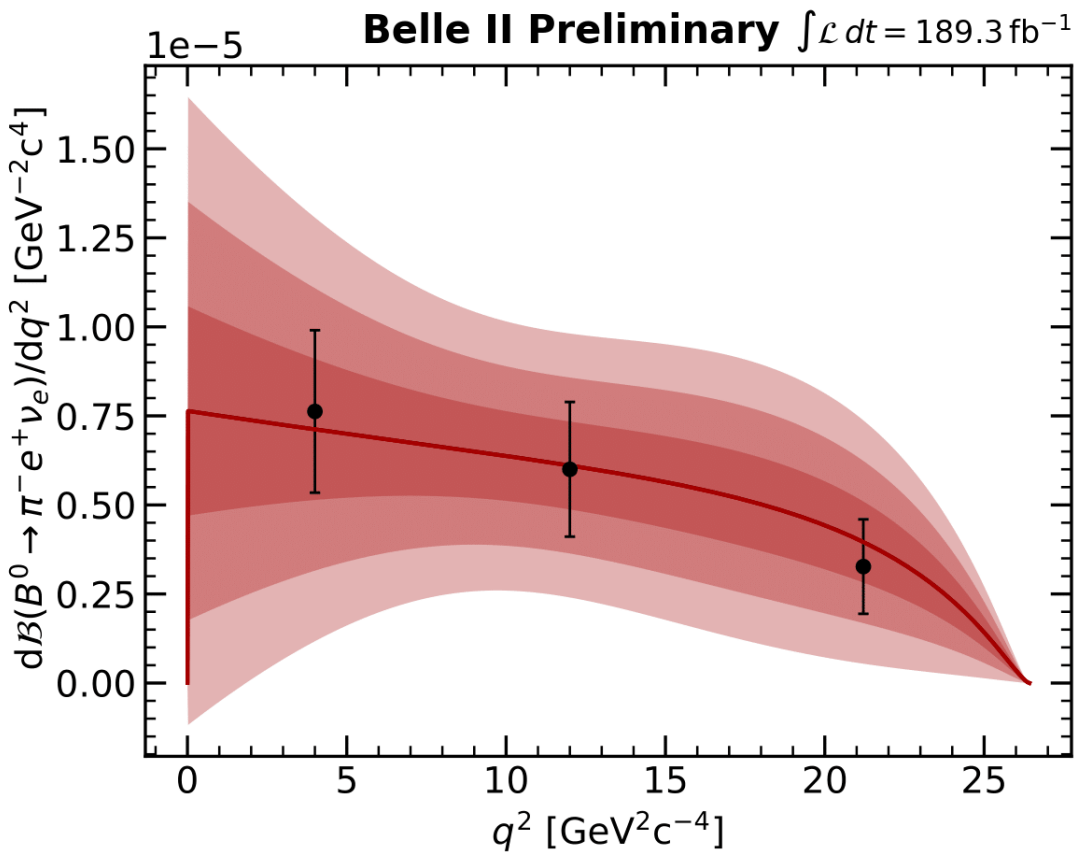}
             \caption{Partial branching fraction uncertainty as a function of $q^2$. The red bands represent 1-3$\sigma$ uncertainties.}
         \end{subfigure}
            \caption{A subset of results for $B^0 \rightarrow \pi^-e^+\nu_e$ from the tagged $B\rightarrow \pi e^+\nu_e$ analysis \cite{tagged_pi_analysis}.}
            \label{fig:nadiaResults}
    \end{figure}

For this study, the tag $B$ meson is reconstructed hadronically using the FEI while the signal $B$ meson is reconstructed in both the neutral $B^0\rightarrow\pi^-e^+\nu_e$ and charged $B^+\rightarrow\pi^0e^+\nu_e$ modes. The signal yield is extracted using an unbinned maximum likelihood fit to $M_\mathrm{miss}^2 \equiv p_\mathrm{miss}^2 = (p_{B_\mathrm{sig}}-p_Y)^2$ where $p_{B_\mathrm{sig}}$ and $p_Y$ are the 4-momenta of the signal $B$ meson and electron-pion ($Y$) system respectively. The fit was performed across the signal region defined by $M_\mathrm{miss}^2 \in [-1, 3]$ GeV$^2$/c$^4$ in three bins of $q^2$ which represents the square of the momentum transfer to the leptonic system. Due to the limited sample size, the various background contributions were combined into a single component for the fit. Distributions of the fitted $M_\mathrm{miss}^2$ for the first $q^2$ bin is shown in Figure \ref{fig:nadiaResults}. Using the fitted yields, partial branching fractions for each $q^2$ bin were calculated and summed for the neutral and charged modes to yield $\mathcal{B}(B^0\rightarrow \pi^- e^+\nu_e)=(1.43\pm 0.27_\mathrm{stat}$ $\pm$ $0.07_\mathrm{sys}$)$\times 10^{-4}$ and $\mathcal{B}(B^+\rightarrow \pi^0 e^+\nu_e)=(8.33\pm 1.67_\mathrm{stat}$ $\pm$ $0.55_\mathrm{sys}$)$\times 10^{-5}$. These are in agreement with the PDG values \cite{pdg}. To extract a value for $|V_{ub}|$, a simultaneous $\chi^2$ fit to the partial branching fractions and the Fermilab/MILC lattice QCD constraints \cite{fnal} was performed. For the fit, the BCL parametrisation \cite{BCL} of the form factors were used. The final value obtained for $|V_{ub}|$ using all $q^2$ bins and both the neutral and charged modes is 
    $$
    |V_{ub}|=(3.88\pm 0.45)\times 10^{-3}
    $$

where the uncertainty above includes both systematic and statistical uncertainties. This study provides the \underline{first measurement of $V_{ub}$ from Belle II} and it is in agreement with the world average \cite{pdg}. For both $B$ modes, the statistical uncertainty dominates so an updated measurement using the entirety of the available Belle II dataset (428 fb$^{-1}$) is in progress. For the $B^+\rightarrow\pi^0e^+\nu_e$ decay, the primary systematic uncertainty comes from the $\pi^0$ efficiency, while for $B^0\rightarrow\pi^-e^+\nu_e$, the FEI calibration factor is the largest systematic. Further details on the analysis can be found in \cite{tagged_pi_analysis}.

\subsection{Untagged \texorpdfstring{$\boldsymbol{B^0\rightarrow\pi^-\ell^+\nu_\ell}$}{B0 → π−ℓ+νℓ}}
\label{SS:22}

\begin{figure}[h]
     \centering
     \begin{subfigure}{0.49\textwidth}
         \centering
         \includegraphics[scale=0.28]{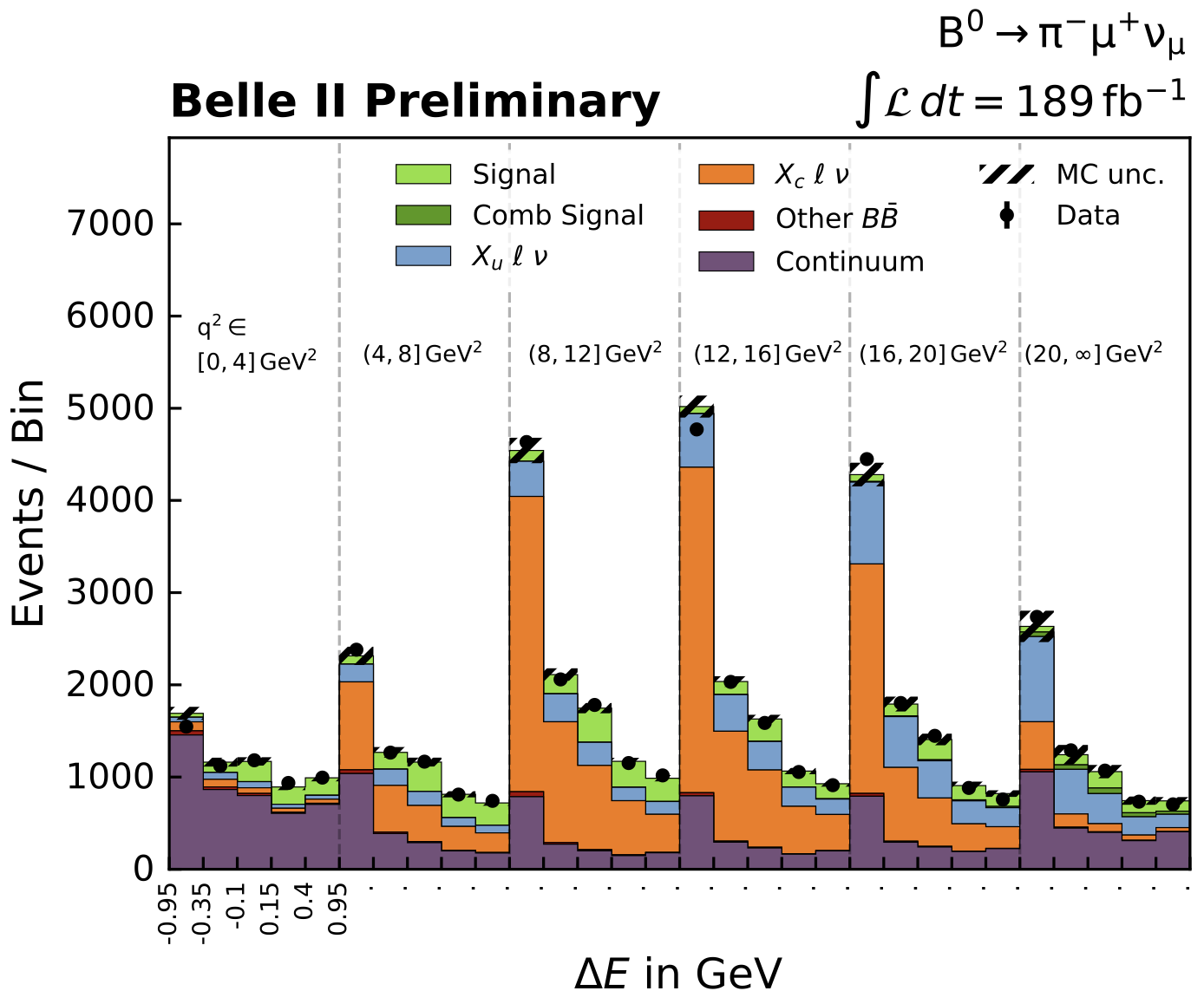}
         \caption{Fit projection of $\Delta E$}
     \end{subfigure}
     \begin{subfigure}{0.49\textwidth}
         \centering
         \includegraphics[scale=0.28]{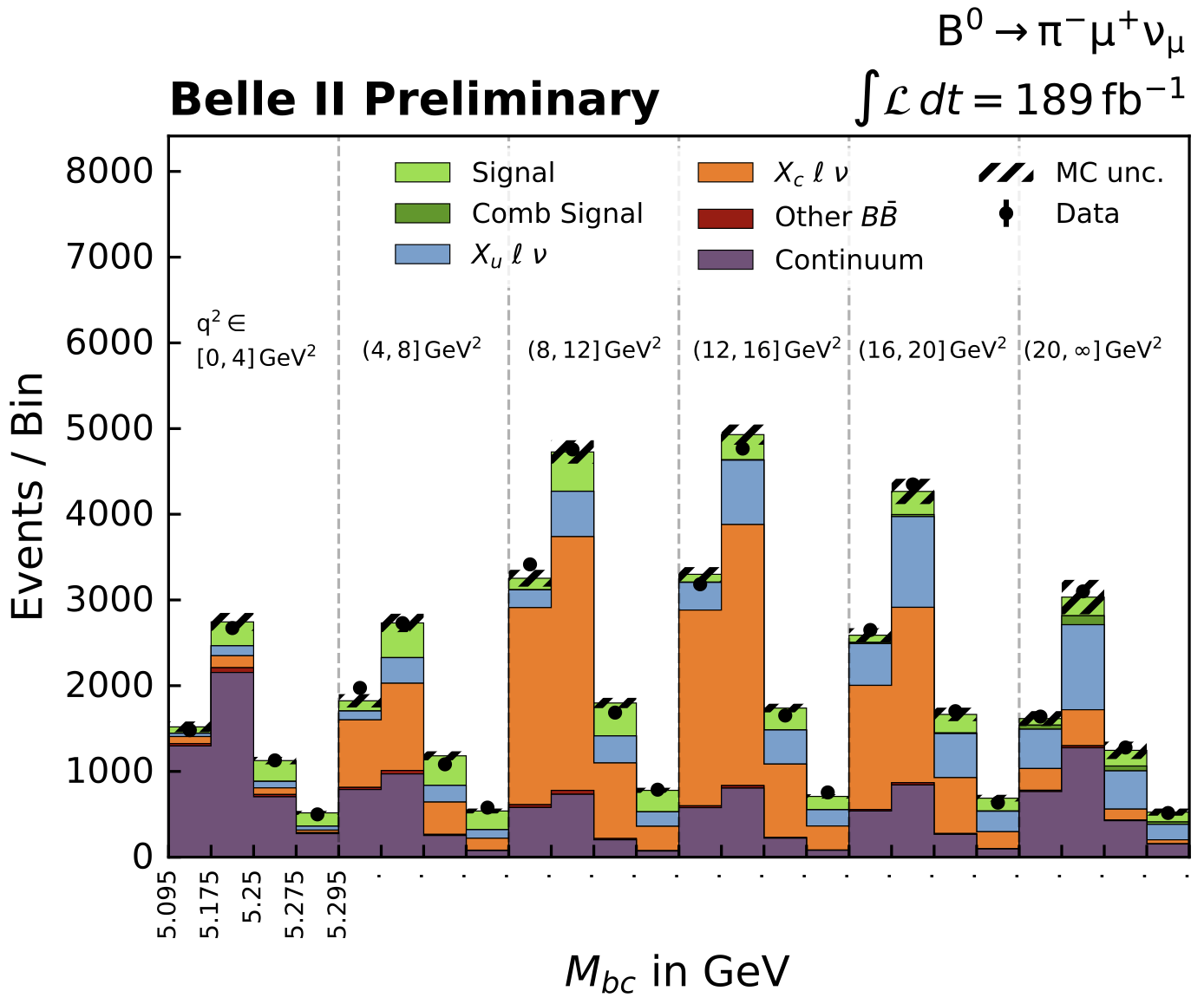}
         \caption{Fit project of $M_\mathrm{bc}$}
     \end{subfigure}
     \\
     \caption{A subset of results for $B^0\rightarrow\pi^-\mu^+\nu_\mu$ from the untagged $B^0\rightarrow \pi^-\ell^+\nu_\ell$ analysis \cite{pi_untagged}. The \textit{Comb Signal} legend refers to combinatorial signal events. \label{fig:resultsSvenja}}
    \end{figure}

Being an untagged analysis, only the signal side $B$ meson is reconstructed as $B^0\rightarrow\pi^-\ell^+\nu_\ell$ with $\ell=e$, $\mu$. The signal is extracted using a 2-dimensional fit to the beam-constrained mass $M_\mathrm{bc} = \sqrt{E^{*2}_\mathrm{beam}-p^{*2}_B}$ and $\Delta E = E^*_B - E^*_\mathrm{beam}$ where $E_B^*$ and $p_B^*$ are the energy and momentum of the signal $B$ meson in the centre of mass frame, and $E_\mathrm{beam}^*$ is equivalent to half of the centre of mass energy. The signal region is defined by $M_\mathrm{bc}\in [5.095, 5.295]$ GeV and $\Delta E\in [-0.95, 0.95]$ GeV. Due to the limited kinematic and event-shape information that accompanies an untagged reconstruction, the signal region is contaminated with large amounts of continuum ($e^+e^-\rightarrow q\bar{q}$ where $q=u,d,s,c$) and peaking background from other semileptonic $B$ meson decays. To suppress such background, multiple boosted decision tree classifiers were used to target and suppress each background component. Following this, a 2-dimensional extended likelihood fit is performed using six bins of $q^2 = (p_B-p_\pi)^2$ where $p_B$ and $p_\pi$ are the 4-momenta of the signal $B$ meson and pion respectively. The number of $q^2$ bins used is double that of the tagged analysis due to the higher statistics that is afforded by the untagged approach. Fit projections for the $B^0\rightarrow \pi^-\mu^+\nu_\mu$ mode is given in Figure \ref{fig:resultsSvenja}. Combining the partial branching fractions calculated from the fitted yields gives $\mathcal{B}(B^0\rightarrow\pi^-\ell^+\nu_\ell)=(1.426 \pm 0.056_\mathrm{stat} \pm 0.125_\mathrm{sys})\times 10^{-4}$ which agrees with the PDG value \cite{pdg}. 

The value of $|V_{ub}|$ is extracted using $\chi^2$ fits to the partial branching fractions for the $q^2$ bins, with Fermilab/MILC lattice QCD constraints \cite{fnal} on the BCL parameters \cite{BCL} included as nuisance parameters. The combined result from using the averaged partial branching fractions from both $\ell=e$, $\mu$ modes is 

     $$
     |V_{ub}| = (3.54\pm0.12_\mathrm{stat}\pm0.15_\mathrm{sys}\pm0.16_\mathrm{th})\times10^{-3}
     $$

which is in agreement with the world average \cite{pdg}. Despite the result being determined from a subset of the available Belle II data, the uncertainty is already competitive with the world average. The dominating systematic uncertainty originates from the modelling of the continuum background due to the limited amount of off-resonance data collected by Belle II so far. Regarding the theoretical uncertainty, this is dominated by the $B\rightarrow\rho\ell\nu_\ell$ form factor. Additional details on the analysis can be found in \cite{pi_untagged}. 

\section{\texorpdfstring{$\boldsymbol{V_{cb}}$}{Vcb} Result}
\label{S:3}

For extracting a value for $|V_{cb}|$, $B$ decays involving a $b \rightarrow c$ transition are needed. The inclusive  $B\rightarrow X_c\ell\nu_\ell$ mode can be used, as well as the $B\rightarrow D^*\ell\nu_\ell$ and $B\rightarrow D\ell\nu_\ell$ modes. Here, the untagged $B\rightarrow D\ell\nu_\ell$ analysis where $\ell=e$, $\mu$ will be presented, with further details found in \cite{Vcb_analysis}. 

\begin{figure}[h]
     \centering
     \begin{subfigure}[t]{0.49\textwidth}
         \centering
         \includegraphics[scale=0.3]{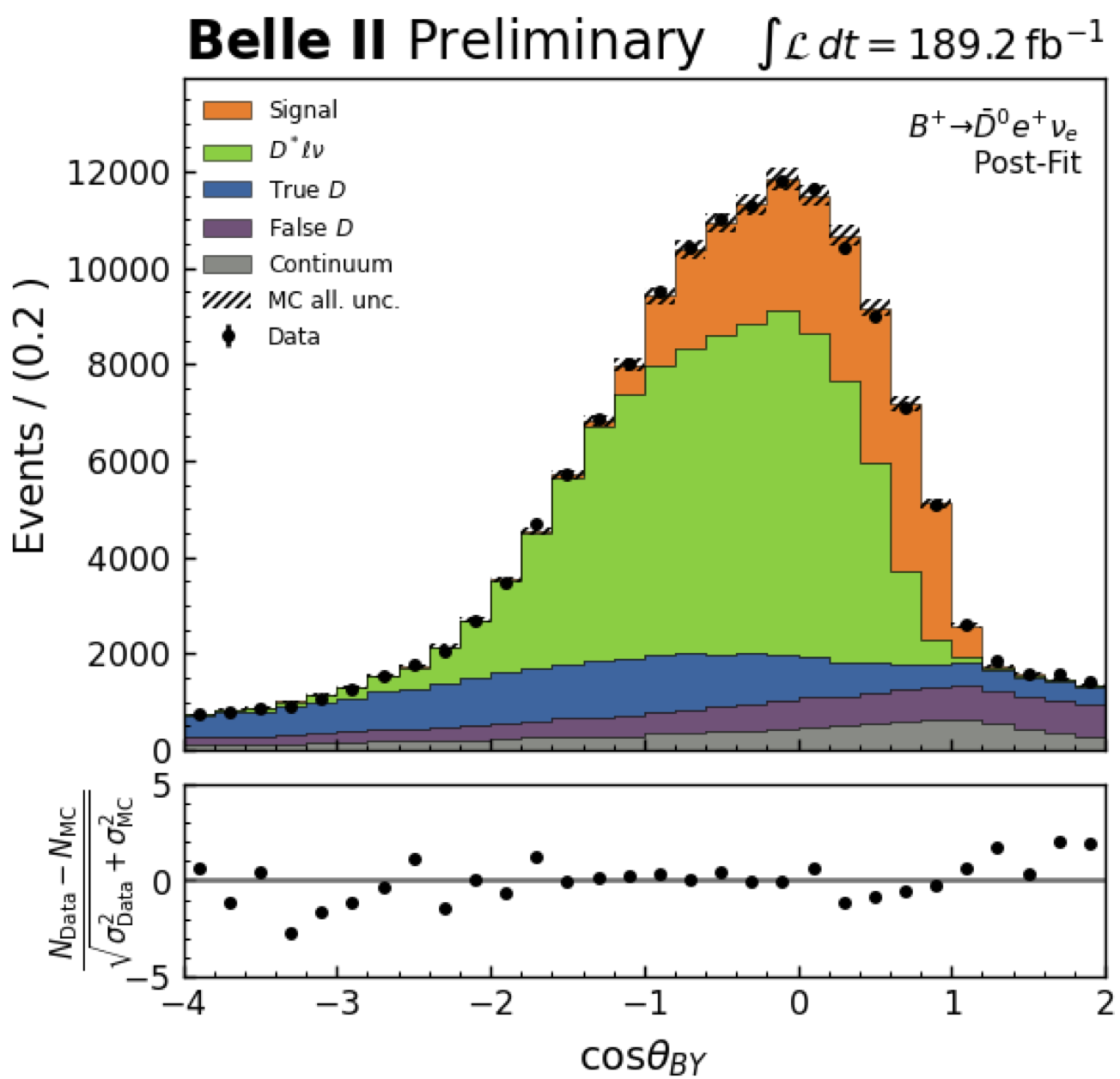}
         \caption{Fitted $\cos\theta_{BY}$ distribution}
     \end{subfigure}
     \begin{subfigure}[t]{0.49\textwidth}
         \centering
         \includegraphics[scale=0.3]{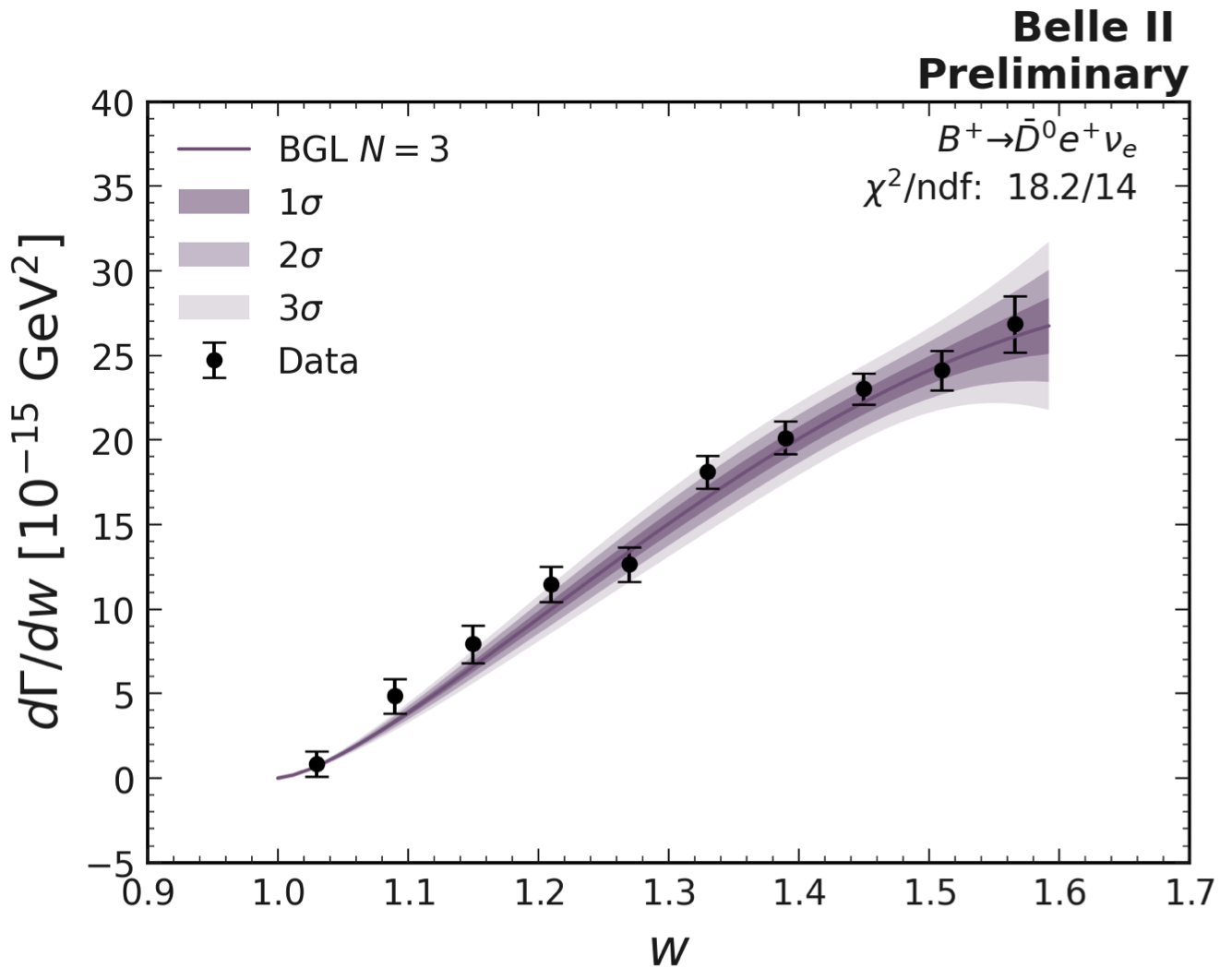}
         \caption{Partial branching fraction uncertainty as a function of $w$}
     \end{subfigure}
     \caption{A subset of results for $B^+\rightarrow \overline{D}{}^0e^+\nu_e$. In sub-figure (a), the \textit{True D} legend refers to events where a $D$ candidate has been correctly reconstructed but has not originated from a $B\rightarrow D\ell\nu_\ell$ event while \textit{Fake D} represents signal events where the $D$ has been incorrectly reconstructed. \label{fig:vcb}}
    \end{figure}

For this analysis, the signal $B$ meson is reconstructed in both the neutral mode $B^0\rightarrow D^-\ell^+\nu_\ell$ ($D^- \rightarrow K^+\pi^-\pi^-$) and charged mode $B^+\rightarrow \overline{D}{}^0 \ell^+\nu$ ($D^0\rightarrow K^-\pi^+$). A significant background component for the charged mode comes from $B^0\rightarrow D^{*-}\ell^+\nu_\ell$ decays. Such background is suppressed by reconstructing an additional slow pion in the event and combining it with the reconstructed $D^0$ candidate, with the event vetoed if the mass difference $\Delta M = M(K^-\pi^+\pi^+)-M(K^-\pi^+)$ is in the range [0.144, 0.148] GeV. The signal is extracted using a binned maximum likelihood fit to the $\cos\theta_{BY} = (2E_B^*E_Y^*-m_B^2-m_Y^2)/(2|p_B^*||p_Y^*|)$ distribution where $E_B^*$, $E_Y^*$ and $p_B^*$, $p_Y^*$ are the centre of mass energies and momenta of the signal $B$ meson and $D, \ell$ ($Y$) system, and $m_B$ and $m_Y$ are their invariant masses. The fit is done across the signal region defined by $\cos\theta_{BY}\in [-4, 2]$ in 10 bins of $w = (m_B^2+m_D^2-q^2)/(2m_Bm_D)$ which is related to the momentum transfer to the lepton-neutrino system $q^2$, and where $m_B$ and $m_D$ are the invariant masses of the signal $B$ and $D$ mesons. The upper bound of $w$ is $\approx$ 1.59 and represents the case where all the $B$ meson energy is transferred to the $D$ meson. A fitted distribution of $\cos\theta_{BY}$ is provided in Figure \ref{fig:vcb} for the $B^+\rightarrow \overline{D}{}^0e^+\nu_e$ mode. To extract a value for $|V_{cb}|$, a combined $\chi^2$ fit is performed to the partial branching fractions and BGL parameterised form factors (with $N=3$) \cite{BGL}, which were calculated with the Fermilab/MILC, HPQCD lattice QCD constraints \cite{HPQCD}. The value of $|V_{cb}|$ extracted is

    $$
    |V_{cb}|=(38.28\pm1.16)\times 10^{-3}
    $$

where the uncertainty includes the statistical, systematic and theoretical components. This value is in agreement with the world average \cite{pdg}. The dominant uncertainty is systematic and arises from the modeling of the background component that consists of incorrectly reconstructed $D$ candidates. 

\section{Test of Lepton Flavour Universality}
\label{S:4}

A test of lepton flavour universality that involves light leptons is given by measuring the following ratio $R(X_{e/\mu}) = \mathcal{B}(B\rightarrow Xe\nu_e)/\mathcal{B}(B\rightarrow X\mu\nu_\mu)$. This study is part of a larger analysis that will measure $R(X_{\tau/\ell})=\mathcal{B}(B\rightarrow X\tau\nu_\tau)/\mathcal{B}(B\rightarrow X\ell\nu_\ell)$ where $\ell=e$, $\mu$. More details on the $R(X_{e/\mu})$ analysis can be found in \cite{ichep}, with a PRL journal paper to be submitted in 2023. For this analysis, the tag $B$ meson is reconstructed hadronically using the FEI, and the signal $B$ meson is reconstructed inclusively in both the neutral $B^0\rightarrow X\ell^+\nu_\ell$ and charged $B^+\rightarrow X\ell^+\nu_\ell$ modes. The variable used to extract the signal yield is the momentum of the lepton in the signal $B$ meson rest frame $p_\ell^*$ which can be determined by using the recoil momentum of the tag $B$ meson. The signal region chosen was $p_\ell^* > $ 1.3 GeV in order to eliminate large amounts of background from events where the signal lepton was incorrectly identified, or when it stemmed from a secondary process such semileptonic $D$ meson decays. A simultaneous binned maximum likelihood fit was performed using fit templates for signal events, continuum background, and background from a mixture of events that were constrained using a sideband of incorrectly charged ($B^{0+}\rightarrow X\ell^-\nu_\ell$) events. The measurement of the $R(X_{e/\mu})$ ratio in the region $p_\ell^* > $ 1.3 GeV is 1.033 $\pm$ 0.010$_\mathrm{stat}$ $\pm$ 0.020 $_\mathrm{sys}$ which is compatible with the Standard Model prediction of 1.006 $\pm$ 0.001 within 1.2$\sigma$ \cite{ichep}. At the time of presenting, this result was the most precise branching-fraction based test of lepton flavour universality, now superseded by the latest $R(K)$, $R(K^*)$ measurement from LHCb \cite{LHCb}. Importantly, the $R(X_{e/\mu})$ result is compatible with the 2019 exclusive result from Belle on $R(D^*_{e/\mu})$ which was measured to be 1.01 $\pm$ 0.01$_\mathrm{stat}$ $\pm$ 0.03$_\mathrm{sys}$ \cite{LFU_belle}. The statistical uncertainty is the second highest uncertainty for the $R(X_{e/\mu})$ measurement, with the highest being the lepton particle identification efficiencies. 

    \begin{figure}[h]
     \centering
     \begin{subfigure}[t]{0.49\textwidth}
         \centering
         \includegraphics[scale=0.35]{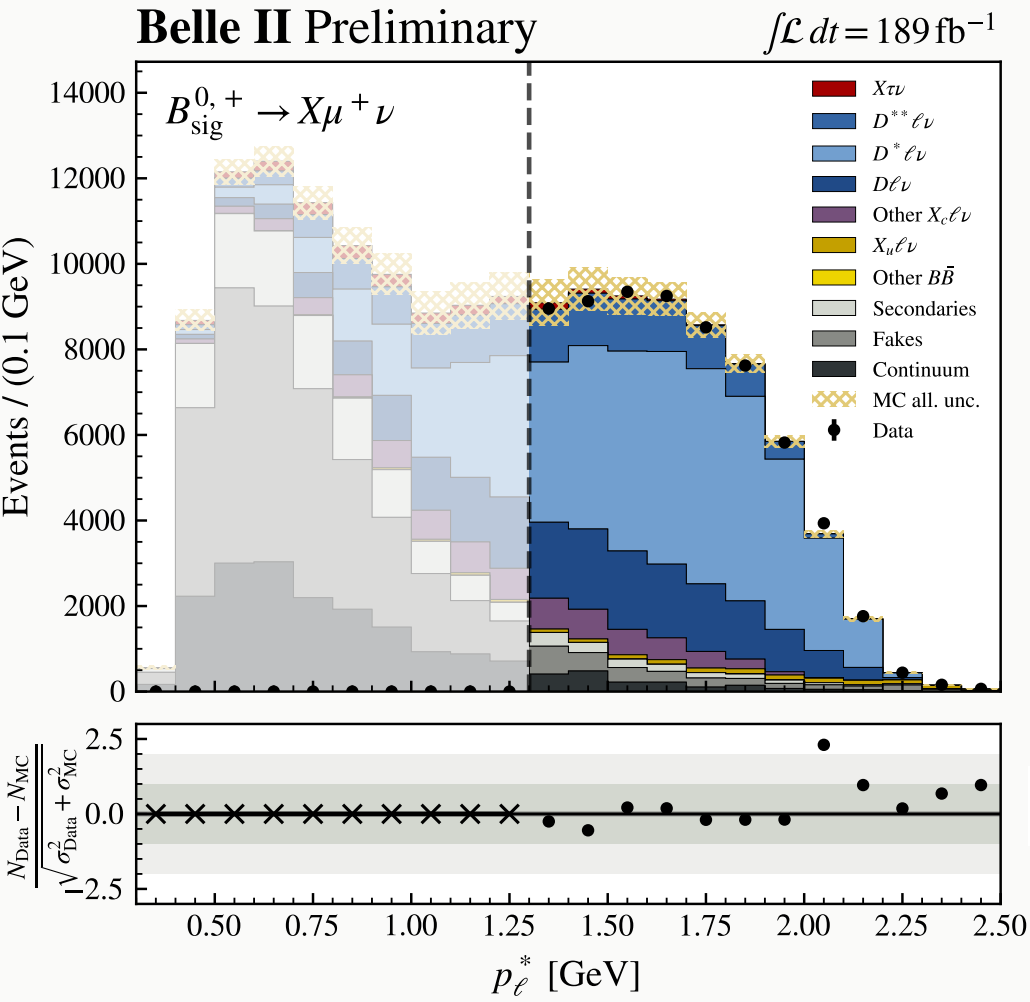}
         \caption{Correct charge events}
     \end{subfigure}
     \begin{subfigure}[t]{0.49\textwidth}
         \centering
         \includegraphics[scale=0.345]{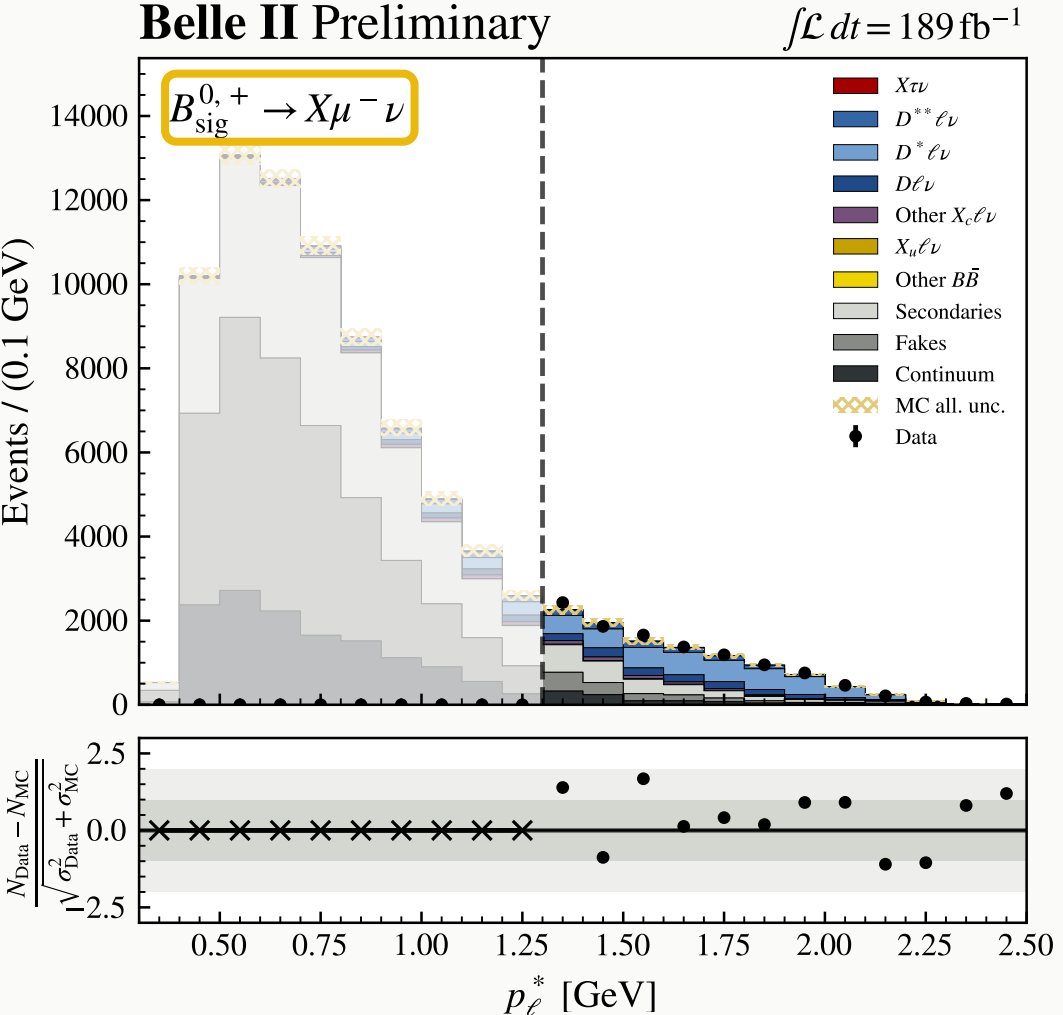}
         \caption{Incorrect charge sideband}
     \end{subfigure}
     \caption{Distributions of $p_\ell^*$ for both correctly and incorrectly charged reconstructed events from the $R(X_{e/\mu})$ analysis \cite{ichep}. The greyed section denotes the region of $p_\ell^*$ that lies outside the signal region. \label{fig:lfu}}
    \end{figure}

\section{Conclusions}

Belle II has so far published results for $V_{ub}$, $V_{cb}$ and $R(X_{e/\mu})$ using 189 fb$^{-1}$ of data, with ongoing efforts to provide updated results for the entire 428 fb$^{-1}$ of data collected so far. Belle II soon plans to publish other results such as the $R(D^*)$ measurement and further measurements of $V_{ub}$ and $V_{cb}$ using both an inclusive and exclusive approach. Although no leptonic $B$ physics results have been published yet, current studies include searches for $B^+\rightarrow \mu^+\nu_\mu$ and $B^0\rightarrow \tau^-\ell^+$, and measuring the branching fraction for $B^+\rightarrow \tau^+\nu_\tau$. With Belle II set to resume operation in 2023, and further upgrades planned for future long shutdowns, Belle II will be able to continue its physics program with strong results over the next decade. 

\bibliographystyle{JHEP} 
\bibliography{main.bib}


\end{document}